\documentclass[preprint,aps]{revtex4}

\usepackage{graphicx}

\begin{document}

\newcommand{\beq}{\begin{equation}}
\newcommand{\eeq}{\end{equation}}
\newcommand{\ben}{\begin{eqnarray}}
\newcommand{\een}{\end{eqnarray}}
\newcommand{\bea}{\begin{array}}
\newcommand{\eea}{\end{array}}
\newcommand{\om}{(\omega )}
\newcommand{\bef}{\begin{figure}}
\newcommand{\eef}{\end{figure}}
\newcommand{\leg}[1]{\caption{\protect\rm{\protect\footnotesize{#1}}}} 

\newcommand{\ew}[1]{\langle{#1}\rangle}
\newcommand{\be}[1]{\mid\!{#1}\!\mid}
\newcommand{\mat}[3]{\langle{#1}\be{#2} {#3}\rangle}
\newcommand{\no}{\nonumber}
\newcommand{\etal}{{\em et~al }}
\newcommand{\geff}{g_{\mbox{\it{\scriptsize{eff}}}}} 
\newcommand{\la}{\lambda}
\newcommand{\lab}{\la\!\!\!\!{\makebox[1pt]{}^-}}
\newcommand{\Ds}{\Delta_s}
\newcommand{\dd}{\partial}
\newcommand{\rr}{{\bf r}}
\newcommand{\Out}{{\rm out}}
\newcommand{\In}{{\rm in}}
\newcommand{\dir}{\hat{\Omega}}
\newcommand{\da}[1]{{#1}^\dagger}
\newcommand{\cf}{{\it cf.\/}\ }
\newcommand{\ie}{{\it i.e.\/}\ }
\newcommand{\paragraphe}[1]{\vrule height1.5pt depth0ex width0pt\paragraph{#1}\vrule
height0pt depth1.5ex width0pt\hfil\break}
\newcommand{\premierparagraphe}[1]{\paragraph{#1}\vrule height0pt depth1.5ex width0pt\hfil\break}
\newdimen\mahauteur
\newdimen\malargeur
\newbox\maboite
\def\jeposecentre#1{\setbox\maboite=#1\mahauteur=\dp\maboite\malargeur=\wd\maboite\divide\malargeur 
by 2\kern-\malargeur\raise\mahauteur#1}
\def\moitie{\kern.1em\raise.5ex\hbox{\the\scriptfont0 1}\kern-0.15em/
  \kern-0.1em\lower.5ex\hbox{\the\scriptfont0 2}}

\title{
%Cavity Quantum Electrodynamics \\ in a wide aperture spherical resonator\\
%Part I : Cavity-induced damping and level shifts
Cavity-induced damping and level shifts \\ in a wide aperture spherical resonator.}

\author{Jean-Marc Daul and Philippe Grangier} 
\affiliation{Laboratoire Charles Fabry de l'Institut  d'Optique, \\ 
 F-91403 Orsay cedex - France}

\begin{abstract}

We calculate explicitly the space dependence of the radiative relaxation rates and  
associated level shifts for a dipole placed in the vicinity of the center 
of a spherical cavity with a large numerical aperture
and a relatively low finesse. In particular, we give simple and useful analytic formulas
for these quantities, that can be used with arbitrary mirrors transmissions. 
The vacuum field in the vicinity of the 
center of the cavity is actually equivalent to the one obtained in a microcavity, 
and this scheme allows one to predict significant cavity QED effects.

\end{abstract}

%\pacs{PACS numbers:  }

\maketitle

\section{Introduction}

Many theoretical and experimental work has been devoted during recent 
years to the so-called ``cavity QED" regime, where strong coupling is achieved 
between a few atoms and a field mode contained inside a microwave or optical 
cavity. In particular, it has been demonstrated that the spontaneous emission 
rate of an atom inside the cavity is different from its value in free 
space \cite{note,D,K,GRGH,GD,HHK,HCTF,JAHMMH,DIJM,ZLML,HF,MYM}. This 
effect can be discussed from several different approaches, and here it will be 
basically attributed to a change of the spectral density of the modes of the 
vacuum electromagnetic field, which is due to the cavity resonating structure. 
This approach is particularly convenient when the cavity does not have one single 
high-finesse mode, but rather many nearly degenerate modes, as it is the 
case in confocal or spherical cavities. More precisely, we will show that a ``wide 
aperture" concentric resonator using spherical mirrors with a large numerical 
aperture, can in principle change significantly the spontaneous emission rate 
even with moderate finesse. Similar result were already demonstrated, using 
either a spherical cavity \cite{HCTF,HF} or ``hour-glass" modes in a confocal cavity 
\cite{MYM}. 
In such experiments, the dipole has to sit within the active region volume, 
which is usually of very small size (of order $(10 \lambda)^3$ to $(100 
\lambda)^3$). In refs \cite{HF,MYM}, a possible solution was implemented by 
reducing the cavity finesse in order to have an 
extended area in which a spherical wave is ``self-imaged" on itself. However, 
getting large effects will put more severe constraints both on the quality of the 
cavity and on the localisation of the dipoles. 

A good understanding of these effects requires first to know the full space 
dependence of the cavity-induced damping and level shifts. In this paper, we 
will look at the situation where an atomic dipole lies close to the center of a spherical 
cavity with a large numerical aperture. We will show that large changes both in 
the atom damping rate and in its energy levels can be expected, even with a 
moderate cavity finesse, provided that the atom sits (relatively, but not 
extremely) close to the cavity center. 
In the following, we will assume that the cavity damping rate $\kappa$ is 
much larger than the free-space atom damping rate $\gamma_{vac}$. In that 
case, the cavity still acts as a continuum with respect to the atomic relaxation,
and the damping rate and level shift of an atom at point $\bf r$ 
are given by \cite{JMD2}:
\ben
\Gamma ({\bf r}) &=& \frac{2\pi}{\hbar^2 } \sum_k ({\bf d.e}_k({\bf r}) )^2 
\delta (\omega_k - \omega _o)   \\
\Delta  ({\bf r}) &=& \sum_k \frac{({\bf d.e}_k({\bf r}) )^2}{\hbar^2}
  {\cal P}\left( \frac{1}{\omega_k-\omega _o } \right)
\een
where the summation are taken over a complete set of modes denoted by the index $k$.
The resonance frequency and field  at point $\bf r$ for mode $k$ 
are respectively denoted $\omega_k$ and ${\bf e}_k({\bf r})$, while the atom
resonance frequency and dipole are respectively $\omega_o$ and ${\bf d} D$, where $D$
is a dimentionless combination of raising and lowering atomic operators.
We note that the free-space value of $\Delta({\bf r})$
is a diverging quantity, which is 
usually assumed to be absorbed in the definition of the atomic levels; therefore, 
one considers here only the (finite)  change of $\Delta({\bf r})$ 
with respect to this free-space 
value, that will be denoted $\Delta'({\bf r})$ :
\beq
\Delta'({\bf r})= \Delta_{cav}({\bf r}) - \Delta_{vac} ({\bf r})
\eeq 

The purpose of this paper is to present an explicit calculation
of $ \Delta'({\bf r}) $ and $\Gamma ({\bf r})/2$, or
equivalently of the modification of the (3-D) vacuum modes spectral density
due to the presence of the cavity.
For definitiveness, we will consider the case of
 an  ``open'' spherical cavity, of radius R and 
of reflectivity and transmittivity coefficients $r$ and $t$, with $r^2 + t^2 = 1$.
The cavity can be ``open'' in the sense that it is made of two separate 
concentric mirrors which do not cover all $4\pi$ steradians.
 We will  assume that $kR \gg 1$ (typically $kR =10^5$ with $k=\omega/c$), and a moderate cavity 
finesse (in the  range 10-100). These parameters seems accessible 
from an experimental point of view, and we will show in the next sections
that they allow one to get quite significant cavity-induced effects.

\section{Modes of a large aperture concentric cavity}

We shall first consider the formal case of a scalar field, before turning to the
real transverse electromagnetic field. Following the ideas of scattering theory,
we propose a computation scheme where the propagation equations are cast in a form that
is suitable for the determination of the mode structure and that allows a convenient
ray-optics formulation.

\subsection{Propagation of a scalar field}

We look for stationnary solutions $\phi(\rr ,t)=\phi(\rr) e^{-i\omega t}$ of
\beq
\Delta \phi - \frac{1}{c^2}\dd^2_{tt} \phi = 0
\eeq
that is, in spherical coordinates $(r,\theta,\phi)$
\beq
\frac{1}{r}\dd^2_{rr}(r\phi) + k^2\phi +\frac{1}{r^2}\Ds \phi =0
\eeq
where we note $k=\omega/c$ and introduce the spherical laplacian
\beq
\Ds = \frac{1}{\sin^2\theta}\dd^2_{\phi\phi}+\dd^2_{\theta\theta}+\frac{\cos\theta}
{\sin\theta}\dd_\theta
\eeq
with eigenvalues $-l(l+1),l\ge0$.

In the far-field regime: $r\to\infty$, we can write unambiguously
\beq
\phi(\rr =r\dir) = \frac{e^{ikr}}{r} f^\Out_r(\dir) + \frac{e^{-ikr}}{r} f^\In_r(\dir)
\eeq
In the following we will often omit the argument $\dir$ of $f^{\Out,\In}$. The quantity
$f^{\Out,\In}_r$ depend slowly on $r$: $\dd_r f^{\Out,\In}_r \sim 
f^{\Out,\In}_r /r$ and tend to large $r$ angular distributions:
\beq
f^{\Out,\In}_r \bea{ccc}\makebox[0pt]{} \\ \longrightarrow \\ \mbox{}^{r\to \infty}
\eea f^{\Out,\In}_\infty
\eeq
Such an `out' field occurs for instance in the case of a radiating localized source:
its squared amplitude then corresponds to the power radiated along $\dir$ per unit
solid angle. Here we also allow for incoming radiation, focused on a localized region
--- which, if not absorbed, turns after focusing into outgoing radiation.
Note that this separation in `in' and `out' field is not possible
too close to the origin, when the Poynting vector is no more almost radial.

\subsubsection{Far-field solution}

We have separate propagation equations for $f^{\Out,\In}_r\,$: the `in' field obeying
\beq \label{eq:f}
\dd_r f_r +\frac{i}{2k}\dd^2_{rr}f_r +\frac{i}{2kr^2}\Ds f_r =0
\eeq
Defining $\delta f = f_r -f_\infty $ we obtain
\beq \label{eq:deltaf}
  \bea[t]{rc} &\dd_r \delta f \\ \rule{0pt}{3ex}\makebox{with orders}& \frac{\delta f}{r} \eea =
 \bea[t]{c} \frac{-i}{2kr^2} \Ds f_\infty \\ \rule{0pt}{3ex}\frac{f_\infty}{kr^2}\eea -
 \bea[t]{c} \frac{i}{2kr^2} \Ds \, \delta f \\ \rule{0pt}{3ex}\frac{\delta f}{kr^2} \eea -
 \bea[t]{c} \frac{i}{2k} \dd^2_{rr}\delta f \\ \rule{0pt}{3ex}\frac{\delta f}{kr^2} \eea
\eeq
(we are only interested in the large $r$ asymptotics, so we first ignore $\Ds$
to get the orders of the different terms).

The first term in the r.h.s. gives the leading behaviour
\beq \delta f \simeq \frac{i}{2kr}\Ds f_\infty \eeq
and a systematic expansion can be obtained by solving iteratively (\ref{eq:deltaf})
producing terms
\beq \label{eq:dlpt}\frac{\Ds^n}{(kr)^{n+p}}f_\infty \; , \left\{ \bea{c}n\ge 1 \\p\ge
0 \eea \right. \eeq

In the sequel, we shall be interested in the field at $\sim 100\lab$ off the origin:
in geometric optics this involves light-rays with an impact parameter smaller than
$100 \lab$, or photons with orbital angular momentum smaller than $100 \hbar$. 
Therefore, in
the multipole expansion of the field, we only keep harmonics with $l\le 100\,$: $\Ds$
is now at most of the order $10^4$. 
Using our typical value  $kR \simeq 10^5$,
the terms (\ref{eq:dlpt}) with $p\ge 1$ are then negligible, and all $p=0$ terms
are obtained by neglecting the last term in (\ref{eq:deltaf}), which is equivalent
to replacing (\ref{eq:f}) with
\beq \dd_r f_r =\frac{-i}{2kr^2} \Ds f_r\eeq
Its solution 
\beq f_r=e^{i\Ds/2kr}\,f_\infty \eeq
then gives all the $p=0$ terms of the expansion (\ref{eq:dlpt}). 
It is easy to show that with $\Ds\sim10^4,\, kr\sim10^5$ 
the magnitude of this first term is a few percent, while
the next two terms range as $10^{-7}$.
Therefore, we shall use only the first term 
$f^\In _r = e^{\frac{i\Ds}{2kr}} f^\In _\infty$  in the following.

To conclude this section, we extend these results to the case of outgoing waves: so far we only
considered incoming radiation, but the analogue of (\ref{eq:f}) is simply given by changing
$i$ to $-i$, and all results are easily transposed under complex conjugation, as an example
of time reversal. In particular, we shall use
\beq f^\Out_r= e^{-\frac{i\Ds}{2kr}}f^\Out_\infty \eeq

\subsubsection{Solution at any distance}

Up to now, we only studied asymptotic expansions of solutions of the wave equation in
spherical coordinates, expressing $f_r(\dir)$ for large $r$ in terms of the values taken by $f_\infty$
in the neighbourhood of $\dir$. We will now derive an integral equation giving $f_r$ at any
finite distance in terms of the function $f_\infty$.

In the case of a stationary wave propagating in free space (with sources at infinity), we expect the
knowledge of $f^\In_\infty$ to enable us to determine the solution $\phi(\rr)$ everywhere,
and in particular its asymptotic behaviour $f^\Out_\infty$: forming wave-packets, this
amounts to constructing the solution at any time, given its value in the infinite past.
We claim that the exact solution is obtained by a sum of plane waves
\beq \label{eq:solua}
\phi(\rr) = -2ik\, \int \frac{d\dir}{4\pi} f^\In_\infty(\dir) e^{i(-k \dir) \rr } \eeq
$f^\In_\infty(\dir)$ being the amplitude of the wave coming from direction $\dir$ with
wave-vector $-k\dir$.

Obviously, the proposed solution does satisfy the wave equation, as a superposition of plane-waves;
to prove our statement it thus suffices to verify that (\ref{eq:solua}) has the right asymptotic
behaviour $f^\In_\infty$. But for large $r$ the integral is dominated by its points of
stationary phase; suppose, for definiteness, that the axis $\theta =0$ is in the direction of $\rr$:
then the phase in (\ref{eq:solua}) $-ikr \cos\theta$ is stationary at $\theta=0$ and $\theta=\pi$.
Near each of these points, the leading contribution to the integral will be of order $1/kr$,
with corrections corresponding to higher powers of $1/kr$: so, the asymptotic part $\phi^{\In,\Out}$
is entirely determined by the leading contribution to the integral. Using

\beq
\int_{{\rm near }\;\theta=0}d\dir\,e^{-iA\cos\theta} 
\bea{ccc}\makebox[0pt]{} \\ \simeq  \\ \mbox{}^{A\to \infty} \eea
\frac{2i\pi}{A}e^{-iA}
\eeq we obtain the contributions of neighbourhoods of $\theta=0$ and $\theta=\pi$ to $\phi(\rr)$
\beq \frac{e^{-ikr}}{r}f_\infty^\In(\theta=0) \; ; \; -\frac{e^{ikr}}{r}f_\infty^\In(\theta=\pi) \eeq
corresponding respectively to the `in' and `out' fields, as could easily have been figured out.

We recognize the right `in'-field in this expansion, and have proved the validity of (\ref{eq:solua}).
Moreover, we have obtained the following relation between `in' and `out' fields
\beq f_\infty^\Out(\dir) = -f_\infty^\In(-\dir) \eeq
If a wave is focused, it emerges in the opposite direction, 
with the opposite phase. 
We also recognize in the $i$ factor in the integral :
\beq \phi(\rr) = \frac{- i}{\la} \int d\dir f^\In_\infty(\dir) e^{i(-k\dir)\rr} \eeq
the relative $\pi/2$ phase at the focus point.

Corrections to this leading behaviour will produce the asymptotic expansion of
$f_r^{\In,\Out}$ in powers of $1/kr$, the neighbourhood of $\theta=0$ contributing to
$f_r^\In$ and that of $\theta=\pi$ to $f_r^\Out$. In this way, one can recover, with longer calculations,
the results of the preceding section. For instance, including the first correction to $f^\In_r$
amounts to replacing $f^\In_\infty(\theta=0)$ with $f^\In_\infty(0)+i/2kr\,\,\big(\Ds f^\In_\infty
\big)(0)$.

\subsubsection{A complete set of explicit solutions}

We know that the angular distribution of the field has variations with $r$ given by an operator
expressed with $\Ds$: thus, eigenfunctions of $\Ds$ - spherical harmonics - give
$r$-independent angular distributions (up to normalization and phase), that is, factorized solutions:
\beq Y_{lm}(\dir) j(r) \eeq
The wave equation and the smoothness of the solution at $r=0$ then determine $j(r)$ up to a 
constant factor:
\beq j(r)= \frac{J_{l+1/2}(kr)}{\sqrt{kr}} \eeq
Using the asymptotics of Bessel functions
\beq \label{dlpt:bessela}
%r\to\infty:\; 
\frac{J_{l+1/2}(kr)}{\sqrt{kr}}
\bea{ccc}\makebox[0pt]{} \\ =  \\ \mbox{}^{r\to \infty} \eea
\sqrt{\frac{2}{\pi}}\frac{1}{kr}\, \sin \Big(
kr -\frac{\pi l}{2}+\frac{l(l+1)}{2kr} \Big) +{\cal O}\big(1/r^3\big) \eeq
we obtain the large $r$ behaviour of this explicit solution
\ben 
\label{eq:explicitea}
\phi({\bf r}= r\dir) &=& -2ik(-i)^l\sqrt{\frac{\pi}{2}}Y_{lm}(\dir) \frac{J_{l+1/2}(kr)}{\sqrt{kr}} \\ \nonumber
&=&
\Big[Y_{lm}(\dir) \frac{ e^{-ikr} }{r} e^{-i\frac{l(l+1)}{2kr}} 
-(-1)^l Y_{lm}(\dir)  \frac{e^{-ikr} }{r} e^{i\frac{l(l+1)}{2kr}} +{\cal O}\big(1/r^2\big) \Big] 
\een
We can verify in this particular case the general relation $f^\Out_\infty(\dir)=-f^\In_\infty
(-\dir)$ and check the action of $\exp i\Ds/2kr$ on $f^\In_\infty$ to the 
accuracy of (\ref{eq:explicitea}).

We can also check the expression for the field at finite
 distance (\ref{eq:solua}): noting that
$f^\In_\infty = Y_{lm}$
%corresponds to \beq \phi = -2ik(-i)^l\sqrt{\frac{\pi}{2}}Y_{lm}\frac{J_{l+1/2}(kr)}{\sqrt{kr}}  \eeq
and choosing normalized spherical harmonics 
\beq \ew {Y_{lm}\mid Y_{lm}} = \int \frac{d\dir}{4\pi} \be{Y_{lm}}^2 =1 \eeq
we decompose
\beq f^\In_\infty = \sum_{lm}\ew{Y_{lm}\mid f^\In_\infty} Y_{lm} \eeq
and obtain
\beq \phi(r\dir) = \sum_{lm}\ew{Y_{lm}\mid f^\In_\infty}\frac{-2ik}{i^l}\sqrt{\frac{\pi}{2}}
Y_{lm}(\dir) \frac{J_{l+1/2}(kr)}{\sqrt{kr}} \eeq
The rotationally invariant quantity $\sum_m \overline{Y_{lm}(\dir)}Y_{lm}(\dir ')$ is conveniently
evaluated when the axis of reference is chosen along $\dir$ and has value $(2l+1)P_l(\cos \alpha)$,
$P_l$ being the $l^{th}$ Legendre polynomial, and $\alpha$ the angle between $\dir$ and $\dir '$.
Using the formula
\beq \sum_{l\ge 0} i^l \frac{J_{l+1/2}(kr)}{\sqrt{kr}} (2l+1) P_l(\cos\alpha) = \sqrt{\frac{2}
{\pi}} e^{i kr \cos\alpha} \eeq
we finally obtain
\beq \phi(\rr) = \ew {2 i k e^{ik\dir ' \rr}\,\mid f^\In_\infty(\dir ') }_{\dir '} \eeq
which reproduces (\ref{eq:solua}).

\subsection{Modes of concentric cavities}

\subsubsection{Perfect spherical resonator}

The explicit solutions given above allow us to determine the field modes inside a
perfectly reflecting sphere, with radius $R \gg \lab$: when we classify modes according
to their spherical symmetry (quantum numbers $l,m$), the requirement that the field shall vanish
on the inner face of the cavity 
\beq f_R=f^\In _R e^{-ikR}+f^\Out _R e^{ikR} = 0 \eeq 
reads
$J_{l+\moitie}(kR)=0$, or according to (\ref{dlpt:bessela}):
\beq \kern3em kR-\frac{\pi}{2}l+{l(l+1)\over 2kR}=0\kern 2em
[{\rm mod}\;\pi] \eeq
for the mode $\kern1em l,m$, 
where we have written all significant terms for $l\sim 100;kR\sim 10^5$, 
obtaining by the way
the lowest order for which the $l$-degeneracy is disproved. 
The eigenfrequencies are then
\beq \nu_{l,n}=\frac{kc}{2\pi}=\frac{c}{2R}\bigg(n+\frac{l}{2}-\frac{l(l+1)}{2\pi kR}\bigg) \eeq
$n$ being the number of radial nodes. 

With our numerical values 
%$R=1{\rm cm}\, ,\la=780 {\rm nm}$ the `free' spectral range $c/2R$ is 15GHz, while 
the $l$-frequency shift has relative magnitude which is up to 
$l/2\pi kR \; \sim 2.10^{-4}$,
and is therefore very small compared to the cavity linewidth. 
%Figure \ref{modes} shows the distribution of these frequency lines, with heights corresponding
%to the relative importance of the modes 
We note that at a point close to the center, small $l$ modes
are more important, since photons travelling close to the origin carry a small orbital momentum.
This point will be made quantitative later, when we will discuss the case of a spherical cavity
with finite transmission.

\subsubsection{Modes in an open cavity}
\premierparagraphe{Principle of the determination of the mode structure}
We recognize the vacuum fluctuations in a concentric cavity as induced by the vacuum
fluctuations of the outer void space which enter into the cavity: we will thus
start our analysis by studying how any incident radiation can enter in the open
concentric resonator.

We consider two spherical mirrors, facing each other in vacuum, with common center~$O$ - as if
in the preceding example the tropical zone of the sphere were transparent, while the polar zones
remained coated. For our computation, we shall replace the infinite vacuum with the inner volume
of a very large sphere centered at $O$, thus replacing a true continuum with a discrete series of very
closely spaced lines. We will use $\cal R$ to denote the radius of the outer closed sphere, and
$R$ for the inner sphere, partially covered by mirrors; and use the following notation for the field
of an eigenmode
\ben {\cal R}>r>R &:&\kern1em \frac{1}{r}e^{ikr}f^\Out_r+\frac{1}{r}e^{-ikr}f^\In_r \nonumber \\
 R>r&:&\kern1em \phi(\rr) \een
Note that the decomposition between `in' and `out' fields in the first equation is allowed
by our choice to study only modes which contribute to $\phi$ near $O$, that is, not too
unfocused. 

We can formally extend $\phi$ for larger values of $r$ as if there were no cavity at all, and
write the far-field decomposition
\beq \frac{1}{r}e^{ikr}g^\Out_r+\frac{1}{r}e^{-ikr}g^\In_r \eeq
$g_\infty^\In$ is the incoming radiation which induces in void space the same field
near $O$ as $f_\infty^\In$ does in the presence of mirrors.
We are ensured that $g^\Out_\infty(\dir)=-g^\In_\infty(-\dir)$ since that field propagates
through the origin. However, the same equality does not hold for $f$: to understand the
relation between $f^\In$ and $f^\Out$, we note that the incoming radiation $f^\In$ induces
a field in the open cavity, which in turn (perhaps after resonance) emits an outgoing field
$f^\Out$; writing the precise relation would require to solve the propagation equation for
$r>R$,
write the boundary conditions on the mirrors, and obtain the condition on $f^{\In,\Out}$
for the existence of a solution $\phi$ between the mirrors satisfying
$g^\Out_\infty(\dir)=-g^\In_\infty(-\dir)$. In some sense, $g$ directly goes through the
origin, while $f^\In$ turns into $f^\Out$ after reflection on the mirrors, or multiple
reflections in the cavity. We will not try to write any explicit formula for that, but use
some general properties of the relation
\beq f^\In_\infty \mapsto f^\Out_\infty =\hat{S}f^\In_\infty \eeq
in close analogy to scattering theory. 
%Here, the operator $\hat S$ is defined independently on $\cal R$ and the outer sphere.
 
We will assume that the losses on the mirrors are negligible when compared to the transmittivity:
the balance between the incoming and outgoing energy fluxes
\beq \| f^\Out_\infty\| ^2 = \| f^\In_\infty\| ^2 =\int {d\dir\over 4\pi}\be{f^\In_\infty}^2 \eeq
requires $\hat S$ to be unitary.

We now express the condition of perfect reflection on the inner face of the large closed cavity, 
i.e. that the field vanishes for $r={\cal R}$ :
\beq %r={\cal R}\Rightarrow {\rm field}=0\: : \kern2em 
e^{ik{\cal R}}f^\Out_{\cal R}+
e^{-ik{\cal R}}f^\In_{\cal R}=0 
\eeq
Once again, we are interested only in those modes that are focused enough so that they may
contribute to the field near $O$: $l$ being bounded, if $\cal R$ is large enough we may use
$f_{\cal R}\simeq f_\infty$ and rewrite the preceding formula
\beq \hat{S}f^\In_\infty=-e^{-2ik{\cal R}}f^\In_\infty \eeq
and formulate the problem of determining the modes of the open cavity (enclosed in a large one)
as follows: find a wavenumber $k$ such that $-e^{-2ik{\cal R}}$ is an eigenvalue of the unitary
operator $\hat S_k$ and identify the corresponding eigenvector. We precised our notation
and used $\hat S_k$ to make clear the dependency of the scattering operator on the frequency
at which the open cavity is excited. However, the problem is not so intricated since
$\hat S_k$ depends slowly on $k$ on the scale of the free spectral range of the $\cal R$-cavity:
\beq \delta k \, R\, {\cal F} \ll 1 \;\Rightarrow\; \hat{S}_k\simeq\hat{S}_{k+\delta k} \eeq
with $\cal F$ the finesse of the $R$-cavity. So, in order to find the eigenmodes with
wavenumbers $k\simeq k_0$ we have to find the eigenvalues and eigenvectors of $\hat S_{k_0}\,$;
then adjust $k$ so that $-e^{-2ik{\cal R}}$ coincides with any chosen eigenvalue of
$\hat S_{k_0}$. Clearly, the mode structure so obtained will be (locally) periodic: the same
eigenvector of $\hat S_{k_0}$ occuring as the far-field of a mode every $\delta k=\pi/{\cal R}$
or $\delta \nu = c/2{\cal R}$.

We would like to stress the close analogy between this general scheme and scattering theory.
The operator 
$\hat S$ is to be thought of as an S-matrix in interaction representation: indeed, we can actually
take the limit of large $r$ for the connection between $f^\In_r$ and $f^\Out_r$, but not
for the relation between $e^{-ikr}f^\In_r$ and $e^{ikr}f^\Out_r$. Following our analogy,
we may say that the radial propagation with phase factor $e^{ikr}$ corresponds to free evolution
in perturbation theory, while the orthoradial propagation of light with changes in the angular
distribution $f$ corresponds to the perturbation and asymptotically vanishes (at large $r$ vs.
at large times in scattering theory).

\paragraphe{Normalization of the modes}
\setbox\maboite=\hbox{$\displaystyle\phi_{\rm real}^2$}
\malargeur=\wd\maboite
\mahauteur = \ht\maboite  %\advance\mahauteur by0.5ex
We define the total energy of the field to be $\int d^3\rr\,\be{\phi(\rr)}^2 = 2\int d^3\rr\,
\strut\copy\maboite\kern-\malargeur\raise\mahauteur\hbox{\vrule height.5ex depth0pt
width0pt\vrule height1.25pt depth-1pt width\malargeur$
_{_{\rm time}} $} $ where $\phi_{\rm real}(\rr,t)={\rm Re}\Big[\phi(\rr)e^{-i\omega t}\Big]$.
For large enough $\cal R$ this energy integral is dominated by large $r$ regions where we can
approximate
\beq \phi(\rr) \simeq {1\over r}e^{ikr}f^\Out_\infty + {1\over r}e^{-ikr}f^\In_\infty \eeq
and so obtain for the energy
\beq {\cal E}= 4\pi \, {\cal R} \, \bigl(\| f^\In_\infty\|^2+\| f^\Out_\infty\|^2\bigr)
= 8\pi \, {\cal R} \, \| f^\In_\infty\|^2 \eeq
We decide to call vacuum the state in which every mode is excited with energy 1 (in real
electrodynamics we should use $\hbar\omega/2$) so that the normalization of any mode in vacuum is
\beq \label{eq:normalisation} \| f^\In_\infty\|^2_{\rm vac}={1\over 8\pi{\cal R}} \eeq
We can easily express the fluctuations of the ordinary vacuum field (here, ordinary means in
infinite space) or rather their spectral density: in a range of frequencies $\delta\nu$ we have
\beq {V\over (2\pi)^3}\int_{\delta\nu}d^3{\bf k}={4\pi V\over c^3}\nu^2\,\delta\nu\eeq
modes for the case of a large volume $V$ with periodic boundary conditions, while
each mode contributes $1/V$ to $\be{\phi(\rr)}^2$ since the energy of any mode is uniformly
distributed in the volume. Finally,one has the expected result in ordinary vacuum:
\beq \kern1em \ew{\be{\phi(\rr)}^2}_{_{\scriptstyle\delta\nu}}=
{4\pi\over c^3}\nu^2\, \delta\nu \eeq
In the next part, we will derive an analogous formula for the case of an open concentric
cavity and compute by how much the vacuum fluctuations (near the center) are amplified or reduced by 
the presence of mirrors at about 1cm. Before that, a last comment is in order: to obtain
the above expression for the ordinary vacuum field we used the fact that the contributions of
different modes add up incoherently. This is always true when we use a basis of modes
in which the energy operator (hamiltonian of the field) is diagonal: in general, eigenmodes
are non-degenerate in frequency and this condition is automatically satisfied. This vanishing
of the off-diagonal matrix elements of the energy reads
\nopagebreak
\beq \int d^3\rr\,\phi_1^*(\rr)\,\phi_2(\rr) =0 \eeq
and expresses the orthogonality of different modes with respect to volume integral.

The discussion of the preceding part enables us to assert a more precise property, which
will be essential in the following: even the angular overlap of different modes is 0.
Indeed, we argued that eigenmodes were obtained by diagonalizing the unitary operator
$\hat S_{k_0}$: but evidently, two different eigenfunctions of the same unitary operator
are orthogonal. Thus, the far-fields of different modes have a vanishing angular overlap,
unless the two particular modes do have the same far-field asymptotics :
%(lines with thesame height on the preceding picture): 
in the latter case, their number of radial nodes
being different ensures the vanishing of their volume overlap.

That property can be stated differently: if we consider all eigenmodes in a range of frequencies
$\delta\nu= c/2{\cal R}$ and associate to every such mode its $f^\In_\infty$ we obtain
a complete orthogonal set of normalized angular functions.
%, normalized according to (\ref{eq:normalisation}).
Should we consider a larger range $\delta\nu$ every member of this orthogonal family would
then be counted $2\,{\cal R}\,\delta\nu\,/c$ times.

\paragraphe{Modification of the vacuum field in an open cavity}
We will find an expression for the field at points close to the origin ($r<$~a few hundreds $\lab$
that is a few dozens $\mu{\rm m}$). All light rays that pass so close to the center will then
reflect almost normally on the mirrors: we will assume that the wave-fronts of all modes are
sufficiently tangent to the mirrors for us to use the $i=0$ reflectivity and transmittivity
coefficients $\rho,\tau$ ($\rho=1$ for a perfect mirror).

Let us first compute the field induced in the open cavity by incident radiation $f^\In_\infty$:
we will note $f^\In_r = U_r f^\In_\infty $ with $U_r\simeq e^{i\Ds/2kr}$ according
to our previous results. A similar relation holds for `out' fields with $\bar{U}_r=U_r^+$ or
$U_r^{-1}$ since $U_r$ is (almost) unitary.
On the outer face of the mirror, the field has incoming and outgoing amplitudes
\beq \begin{array}{l} {e^{-ikR}\over R}\: U_Rf^\In_\infty \\ \\ {e^{ikR}\over R}\:
\bar{U}_Rf^\Out_\infty \end{array}
\eeq
while on the inner face we find
\beq \begin{array}{l} {e^{-ikR}\over R}\: U_Rg^\In_\infty \\ \\ {e^{ikR}\over R} \:
\bar{U}_Rg^\Out_\infty \eea
\eeq
with \beq 
\label{systeme:1}
g^\Out_\infty(\dir)=-g^\In_\infty(-\dir):\kern1em g^\Out_\infty=-\hat{P}g^\In_\infty\eeq
$\hat P$ being the parity operator, that commutes with $\Ds$ and $U_R$. The $g^\In$ wave
has two contributions: partial transmission of $f^\In$ and partial reflection of $g^\Out$:
\beq \label{systeme:2}
e^{-ikR} U_R \; g^\In_\infty(\dir) = -\rho(\dir) e^{ikR} \bar{U}_R \; g^\Out_\infty(\dir) + \tau(\dir)
e^{-ikR} U_R \; f^\In_\infty(\dir) \eeq
where we explicitly write the angular dependency of $\rho,\tau$: in particular, out of the mirrors
$\rho=0$ and $\tau=1$. We can easily solve (\ref{systeme:1},\ref{systeme:2}) to obtain
\beq g^\In_\infty(\dir) = U_R \frac{1}{U_R^2 - e^{2ikR} \rho(\dir) P}\tau(\dir) U_R \cdot f^\In_\infty(\dir) \eeq
which yields a formula for the field at any point through
\beq \phi(\rr) = \ew { 2ike^{ik \dir \rr} \mid g^\In_\infty(\dir) }_{\dir} \eeq
Introducing the shorter notation
\beq {\cal T} = \tau(\dir) {1 \over U_R^2 -e^{2ikR}P \rho(\dir) } U_R \eeq
we obtain for the vacuum field
\beq \ew{\be{\phi(\rr)}^2}_{_{\scriptstyle \rm vac}} = \sum_{{\rm modes:}\; f^\In_\infty} 
\mat{2ike^{ik \dir \rr}}{\bar{\cal T}^+ \: U_R}{f^\In_\infty}\:
\mat{f^\In_\infty}{U_R^+\: \bar{\cal T} }{2ike^{ik \dir \rr}} \eeq
The above-mentioned property of orthogonality of the modes now gives a considerable
simplification since we do not need to know the precise expression of all the modes, but  only
their total contribution to the physically meaningful quantity 
$\ew{\be{\phi(\rr)}^2}_{_{\scriptstyle \rm vac}}$ in a given frequency range: so, no matter
what the true $f^\In_\infty$ may look like, they surely give a closure relation. Recalling that
the $f^\In_\infty$ associated to the modes in a frequency range of $c/2{\cal R}$ form a complete
orthogonal set normalized according to (\ref{eq:normalisation}) we see that
\beq \sum_{{\rm modes\: in}\:\delta\nu :\; f^\In_\infty} \mid f^\In_\infty \rangle
\langle f^\In_\infty \mid = {2{\cal R}\,\delta\nu\over c} {1\over 8\pi{\cal R}} \: {\bf 1} \eeq
and is actually independent of the large cavity we used to mimic the infinite vacuum.
We then have
\beq \ew{\be{\phi(\rr)}^2}_{_{\scriptstyle \rm vac}} = {k^2\,\delta\nu \over \pi c} \| U_R^+
\,\, \bar{\cal T} \cdot e^{ik\dir\rr} \|^2 
\eeq
Using the unitarity of $U_R$ and the expression of ordinary vacuum fluctuations we find
\beq \label{eq:de:base}
{ \ew{\be{\phi(\rr)}^2}_{_{\scriptstyle \rm cav}}\over
\ew{\be{\phi(\rr)}^2}_{_{\scriptstyle \rm vac}} } = \| {\cal T} \cdot
e^{-ik\dir\rr} \|^2 = \| \tau {1 \over e^{i\Ds/kR} -e^{2ikR}P \rho } e^{i\Ds/2kR} \cdot
e^{-ik\dir\rr} \|^2 \eeq

\subsubsection{Numerical study and ray optics interpretation}
\premierparagraphe{Numerical results}
As expressed in (\ref{eq:de:base}), we shall evaluate the action of an operator
on the function $\dir\mapsto e^{-ik\dir\rr} $ and then compute the squared norm
of the resulting function: this can be done numerically, decomposing functions on
spherical harmonics
%\beq e^{-ik\dir\rr} =\sum_{\bea{c} l,m \\ 0\le \be{m}\le l \eea} (-i)^l\,\sqrt{\pi
\beq e^{-ik\dir\rr} =\sum_{l,m; \; 0\le \be{m}\le l} (-i)^l\,\sqrt{\pi
\over2}\,{J_{l+\moitie}(kr)\over\sqrt{kr}}\,\overline{Y_{l,m}\bigl(\rr/r\bigr)}\,
Y_{l,m}\bigl(\dir\bigr) \eeq
and computing matrix elements of the $k\/$-dependent operator in the basis
of spherical harmonics.

We only considered axially-symmetric cavities: in that case, the operators involved 
conserve $m$ and the result is expressed as a sum of contributions from the different
$m$ sectors. Moreover, $m=0$ gives the single contribution to the field on the axis.
In the latter case, we could use truncated systems of up to $l_{\rm max}=300$
spherical harmonics and study the effect of truncation: the result was constant within
1\% for $l_{\rm max}\ge 100$. In the general case (field fluctuations at points
away from the symmetry axis of the cavity), we used $l_{\rm max}=100$ with any $m$
to study the spatial and spectral dependency of vacuum fluctuations.

\begin{figure}[ht]
%\begin{center}  
\includegraphics[width=8cm]{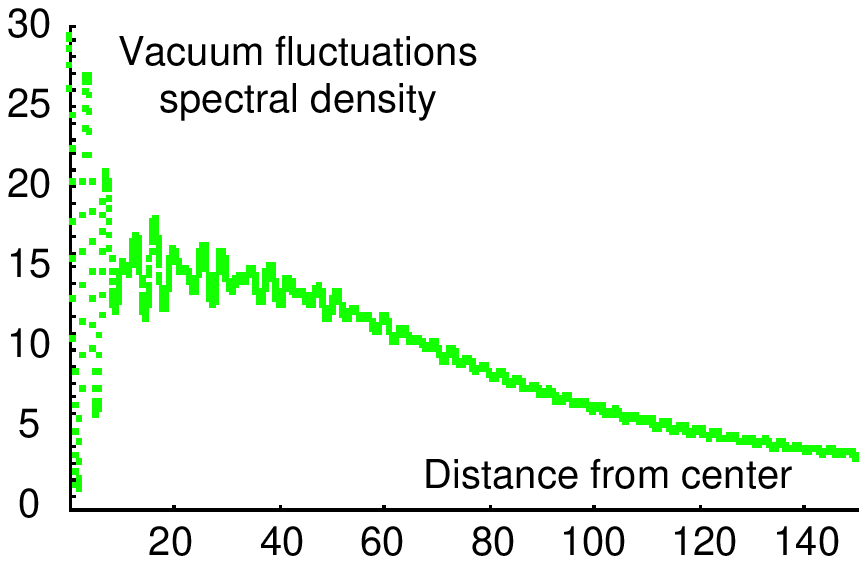} 
\includegraphics[width=8cm]{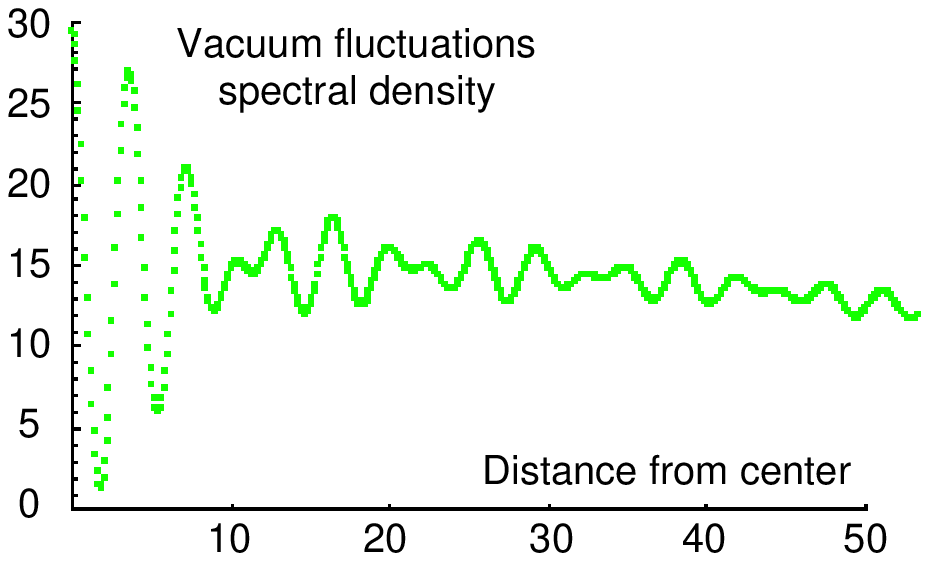} 
%\end{center}
%\vskip 5cm
\caption{ 
 Spatial variation of vacuum fluctuations on the cavity axis, obtained
from a full numerical calculation.
The horizontal axis unit is $1/k$, and the vertical axis unit is the standard vacuum level.
The amplitude reflection coefficient of the mirrors is taken to be $\rho=0.98$,
and the numerical aperture of the cavity is 0.7.
The right hand side of the figure is a zoom close to the cavity center. }
 \label{num}
\end{figure}

The graphs shown on fig. \ref{num} were obtained with concentric mirrors of uniform reflectivity
$\rho=0.98$ (giving an intensity transmittivity $T\simeq 4\%$) covering 30\% of the
$4\pi$ steradian; this is obtained for an half-aperture angle $\theta_m = 45^o$ for each mirror. 
The frequency is set at the resonance value at the cavity center. 
%\centerline{courbes: champ sur l'axe, hors de l'axe: a resonance}
%\centerline{spectre en l'origine, hors de l'origine}
We note the rapid oscillations of the amplitude of vacuum fluctuations near
the center of the cavity; right at the center, the obtained value at resonance
agrees approximately
with the usual rough estimate $4/T \times\,\Delta\Omega_{\rm mirr}/
4\pi$. However, at a few $\mu\/$m away from the center the enhancement effect
is halved, and then decreases further on a scale of $\sim 15 \la$. The next sections will give
support to a qualitative and quantitative formulation of these facts in terms
of ray-optics.

\paragraphe{Case of a closed cavity: ray optics interpretation}
We come back to the case of a closed spherical resonator, now allowing a non-zero
transmittivity of the mirrors (but still negligible losses, as we always suppose in this
article). We can apply to that particular case the formalism we developed for open cavity,
and recover the known modes involving spherical harmonics:
\beq {1\over e^{i\Ds/kR} -e^{2ikR}P \rho } e^{i\Ds/2kR} \,Y_{l,m} =
{1\over e^{-il(l+1)/kR}-(-1)^l\rho e^{2ikR} } e^{-il(l+1)/2kR} Y_{l,m} 
\eeq
so that the vacuum fluctuations (normalized to 1 for usual vacuum) read
\beq \sum_{l\ge0} {T\over \be{e^{-il(l+1)/kR}-(-1)^l\rho e^{2ikR}}^2}\,{\pi\over 2}
\, (2l+1)\,{J_{l+\moitie}(kr)^2\over kr}
\eeq
where we recognize the usual resonance factor, including the slight non-degeneracy
of modes (increasing $l$: lower resonant frequency) and see the explicit spatial-dependency
of $l$-modes.

Averaging vacuum fluctuations in a frequency range much larger than the `free' spectral
interval of the cavity we can replace the frequency-dependent factor by its average value:
1, and use the following sum rule
\beq \sum_{l\ge0} \,{\pi\over 2} \, (2l+1)\,{J_{l+\moitie}(kr)^2\over kr}=1
\eeq
We actually recover usual vacuum if the cavity is large enough (with free spectral range
smaller than the experimentally used frequency bandwidth).

If we are interested in the behaviour of vacuum fluctuations in the vicinity of the center
(a few microns) where only small $l$ modes contribute, we may assume that all modes
are degenerate and use a common resonant factor:
\beq \frac{T}{\be{1-(-1)^l \rho e^{2ikR}}^2} \eeq
The sum rules for Bessel functions
\beq \sum_{l\/{\rm:odd/even}}\,{\pi\over 2} \, (2l+1)\,{J_{l+\moitie}(kr)^2\over kr}=
{1\over2}\pm {\sin\,2kr \over 4kr} \eeq
then allow us to write the vacuum field as
\beq {T\over \be{1- \rho e^{2ikR}}^2 }\biggl({1\over2}+ {\sin\,2kr \over 4kr}\biggr)
+{T\over \be{1 + \rho e^{2ikR}}^2} \biggl({1\over2} - {\sin\,2kr \over 4kr}\biggr)
\eeq
At any point, we have two series of resonant lines, in which the vacuum noise is distributed
with weights $\moitie\pm\sin2kr/4kr$.
%\centerline{graphes correspondants}
At a frequency which is resonant for the center of the cavity, the spatial dependency
of the vacuum field shows a reduction by a factor 2 when one moves away from the center,
as was noted above in the case of an open cavity; we interpret this as the distribution
of vacuum fluctuations on the two series of lines: right at the center only the $l=0$
mode appear, but the odd $l$ modes have no other common nodes and share
$\simeq$50\% of the vacuum noise away from the center. The same conclusions can be
formulated in terms of light-rays: noting that
\beq {1\over2} + {\sin\,2kr \over 4kr} = \int {d\dir\over 4\pi}\;\cos^2 k\dir\rr, \; \; \; \; 
{1\over2}- {\sin\,2kr \over 4kr} = \int {d\dir\over 4\pi}\;\sin^2 k\dir\rr \eeq
we may express the vacuum fluctuations (still neglecting the $l$-dependency of
resonant frequencies) as
\beq
{ \ew{\be{\phi(\rr)}^2}_{_{\scriptstyle \rm cav}}\over
\ew{\be{\phi(\rr)}^2}_{_{\scriptstyle \rm vac}} } = \int {d\dir\over 4\pi} \left(
{T\over \be{1- \rho e^{2ikR}}^2 }\cos^2\bigl(k\dir.\rr) \; +\;
{T\over \be{1+ \rho e^{2ikR}}^2} \sin^2\bigl(k\dir.\rr) \right)
\eeq
and propose the following interpretation: through the center of the cavity we may
draw a ray in any direction; that ray reflects on the inner face of the cavity back
onto itself and gives rise to a system of stationnary waves. For any such ray, the
field oscillations (forced by the outside vacuum) may have maximum amplitude or a node
at the center: correspondingly, the stationnary wave will have squared amplitude
$\cos^2k\dir\rr$ or $\sin^2k\dir\rr$ at the point of interest.

As for the contribution of rays that support a mode having an antinode at the cavity
center, they have different phases away from the origin according to their direction
and thus contribute in the average with weight $\moitie$: the former considerations
on the positions of nodes of Bessel functions are now reformulated as positions
of nodes of stationary waves along rays with different directions.

\begin{figure}[ht]
%\begin{center}  
\includegraphics[width=8cm]{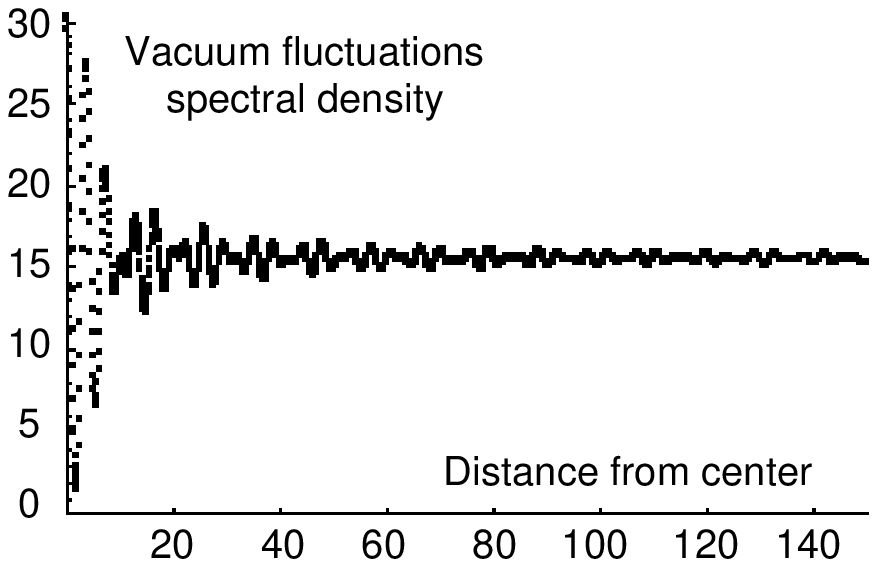} 
\includegraphics[width=8cm]{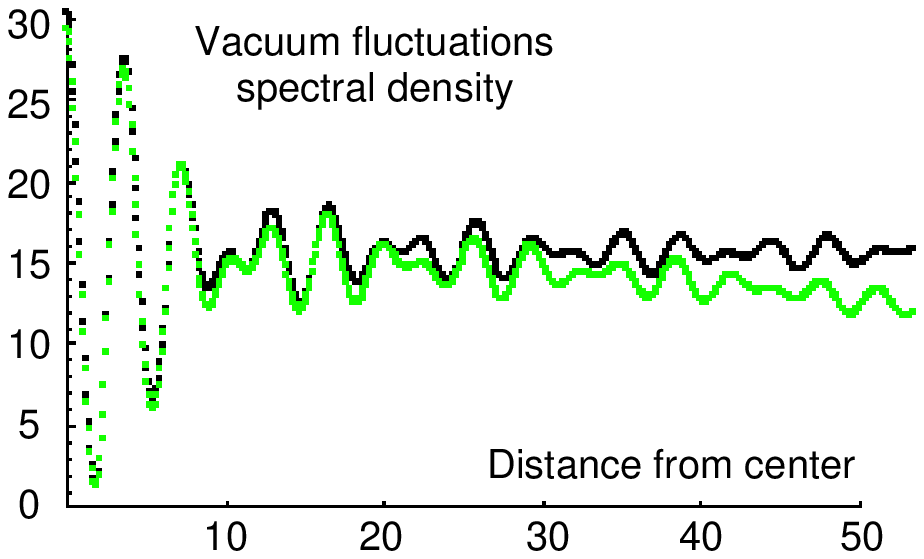} 
%\end{center}
%\vskip 5cm
\caption{ 
Spatial variation of vacuum fluctuations on the cavity axis,
obtained from a simple ray-optics calculation.
The horizontal axis unit is $1/k$, and the vertical axis unit is the standard vacuum level.
The amplitude reflection coefficient of the mirrors is taken to be $\rho=0.98$,
and the numerical aperture of the cavity is 0.7.
The right hand side of the figure is a zoom close to the cavity center,
which shows also a comparison with the full calculation (fig. \ref{num}). }
 \label{compa}
\end{figure}

\paragraphe{Spherical aberrations}
As it can be seen on fig. \ref{compa},
the simple ray-optics analysis described above agrees approximately with the results of the
operator-based numerical computation in the vicinity of the center,
but fails to describe the vacuum fluctuations away from that area. 

Indeed, in order to handle field properties at a point located at a distance $r$
form the origin, we shall not consider anymore the effect of rays going through
 the origin, but rather that of rays going through this point : such rays miss 
the origin by a distance $d < r$, and thus carry $\ell < kr$ orbital momentum
in $\hbar$ units. These rays do not close after one round trip : 
if we do not move too far from the origin, we may still assume that any point is 
imaged onto its symmetric after one reflection, and so onto itself after
two reflections, but the (twice) reflected way is tilted by an angle 
$\delta \theta = \frac{4r}{R} \sin(\theta)$.
However, as we are considering finesse values in the range 10-100, and $r$, $R$
values respectively smaller than $10^2$ and $10^5 \lambdabar$, we may 
reasonably neglect this tilt and and associate to any ray an average value $\theta$.
Note that $\theta$ corresponds to $\ell$ via $\ell = k r \sin(\theta)$. since
$\ell$ is conserved at reflection on the mirror, due to the symmetry with 
respect to a radius, the change in the direction of the light ray is actually
accompanied by a lack of re-imaging of the point back to itself. 
So, our assumption really consists in neglecting both the tilt and the 
lack of re-imaging, and correspondingly in keeping the spatial
modulation $cos^2 k r$ and $sin^2 k r$ in the expressions given above.

What cannot be neglected, however, is the relative phase with which the 
reflected light comes back to the initial point : for off-center rays,
a round trip involves propagation on a distance $ 4 R + \frac{2 r^2}{R} \sin^2(\theta)$. Consequently, the frequency-dependant factor acquire an extra
phase term and becomes  $ 2 i k R + \frac{i k r^2}{R} \sin^2(\theta)$.
As it can be seen on fig. \ref{compb}, this phase term is basically  responsible for the 
decrease of the spectral density as a function of the distance.
%where we have used the relation $\ell = k r \sin(\theta)$ in order to
%recover the $\ell$ dependence of the resonant frequencies in the semiclassical
%(large $\ell$) limit. 

\begin{figure}[ht]
%\begin{center}  
\includegraphics[width=8cm]{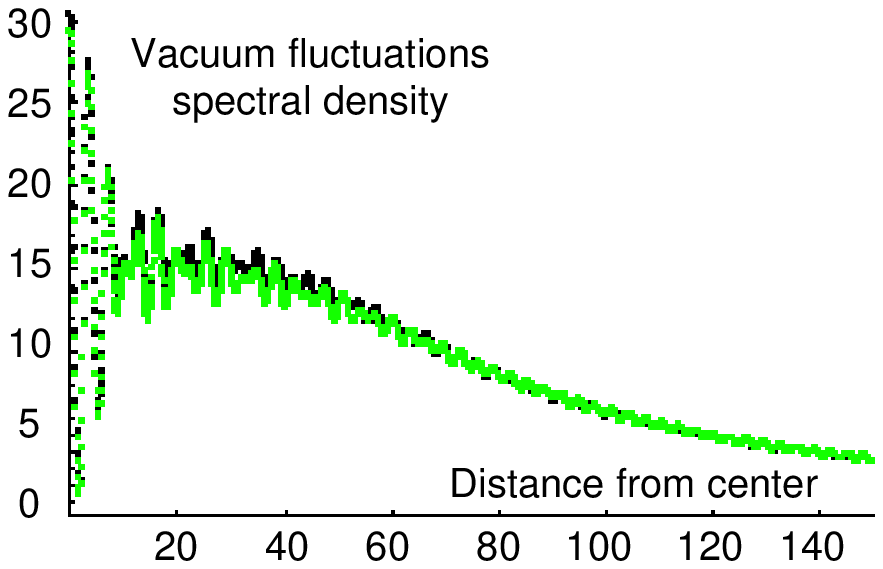} 
\includegraphics[width=8cm]{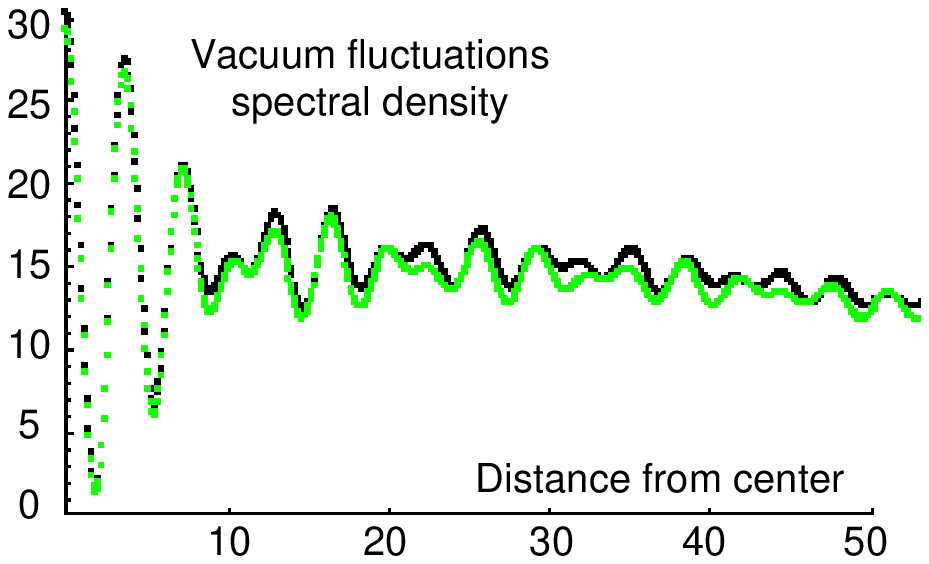} 
%\end{center}
%\vskip 5cm
\caption{ 
Same as fig. \ref{compa}, now including spherical aberrations, so that
the spectral density is now decreasing as a function of the distance.
The result of the full calculation (fig. \ref{num}) is shown for comparison. }
 \label{compb}
\end{figure}

\paragraphe{Boundary effects and diffraction losses}
The formulation in terms of light rays described above can be straightforwardly
extended from a closed to an open cavity : to any ray we associate an enhancement
factor, which is frequency-dependant if the ray meets the mirror, and is unity
if the ray misses the mirror, as well as a spatially-dependant term describing
the intensity modulation of the stationnary waves. 
%\pagebreak

When the two mirrors are identical,
we use the results known for symmetrical Fabry-Perot resonators and obtain :
\beq
\label{rsr}
{ \ew{\be{\phi(\rr)}^2}_{_{\scriptstyle \rm cav}}\over
\ew{\be{\phi(\rr)}^2}_{_{\scriptstyle \rm vac}} } = 
\int {d\dir\over 4\pi} 
({T\over \be{1- \rho e^{2ikR} e^{i \frac{k r^2}{R} \sin^2 \theta}}^2} \cos^2\bigl(k\dir.\rr) \; +\;
{T\over \be{1+ \rho e^{2ikR} e^{i \frac{k r^2}{R} \sin^2 \theta}}^2} \sin^2\bigl(k\dir.\rr) )
\eeq
Should the two mirrors be different, the formula would be easily modified, 
just as in the case of a usual Fabry-Perot resonator (see Appendix A). 
In particular, the maximal enhancement at the center of a symmetrical cavity 
subtending a total solid angle $\Omega_{m}$is :
\beq
{ \ew{\be{\phi(0)}^2}_{_{\scriptstyle \rm cav}}\over
\ew{\be{\phi(0)}^2}_{_{\scriptstyle \rm vac}} } = 
{\Omega_{vac} \over 4\pi} + {T\over \be{1- \rho}^2}{\Omega_{m} \over 4\pi}
\eeq

It is worth noting that the ray-formula without spherical aberrations given above
corresponds exactly to the result of the more rigourous analysis, when one 
neglects $\Delta_S$, i.e., if one takes $e^{i {\Delta_S \over k R}}$ equal to unity.
Though the main effect of $\Delta_S$ is accounted for by spherical
aberrations, a small discrepancy remains : for $\rho_{m} = 0.98$
and ${\Omega_{m} \over 4 \pi} = 0.3$, the numerical computation in the basis
of spherical harmonics yields an enhancement factor of $29.2$ at the cavity center 
and at resonance, while the ray computation gives $30.4$ in the same conditions. 
We shall explain this small difference by diffraction losses : those rays that would
be reflected near the edge of the mirror are actually lost due to diffraction
and fail to do as many round-trips as the other rays. This second effect of
$\Delta_S$ can be estimated by looking for an approximate inverse 
of the operator $(e^{i {\Delta_S \over k R}} - \rho(\theta))$, valid near
the mirror edge. The result of this procedure is that one can still 
use the previous formula for any detuning and at any point, 
provided that the boundary value $\theta_{m}$ is decreased to
$\theta_{eff} = \theta_{m} - \delta \theta$, with 
$\delta \theta = 1/ \sqrt{k R T}$ for symmetrical mirrors, 
and $\delta \theta = 1 / \sqrt{k R (1-\rho_{av}^2)}$ for 
non-symmetrical mirrors, $\rho_{av}$ being the average reflectivity
$(\rho_{1} + \rho_{2})/2$ of the two mirrors.
Applying this procedure to the above example, we shall substract 
\beq {1\over \sqrt{k R T}} \sin(\theta_{m})  ({T\over \be{1- \rho}^2} -1) = 1.2, \eeq
which is quite satisfactory since $30.4-1.2 = 29.2$. 

The comparison of the ray calculation and of the complete one for an open
cavity is shown on fig. \ref{compc}, with the same parameters as for fig. \ref{num}.
As it can be seen, the agreement is very good,
and justifies $a\; posteriori$ the assumptions which have been made. It can therefore
be concluded that the main correction to the naive calculation is
the phase error due to spherical aberrations, with some small
correction from the edge diffraction losses. These corrections
are enough to get the right answer in the conditions that we are considering
($R \sim 10^5\lab$, $r$ smaller than $100 \lab$).

\begin{figure}[ht]
%\begin{center}  
\includegraphics[width=8cm]{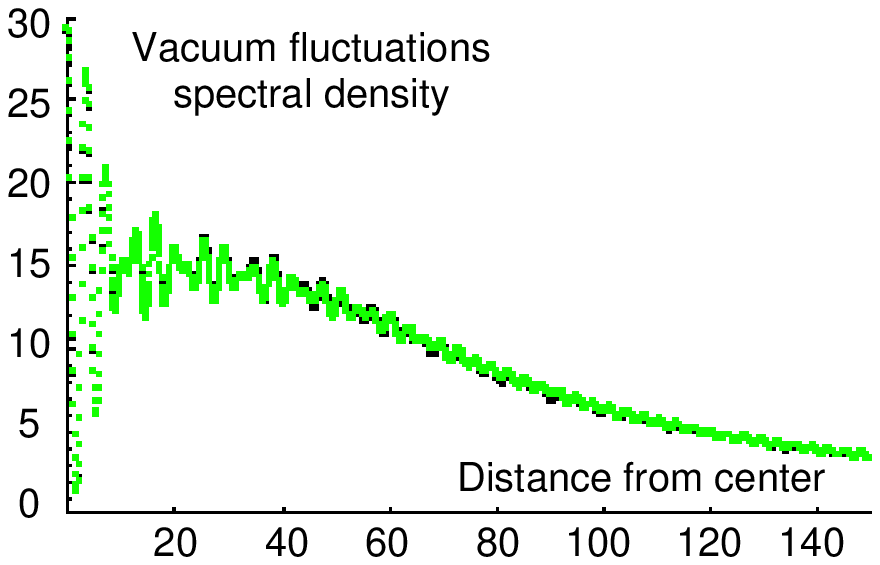} 
\includegraphics[width=8cm]{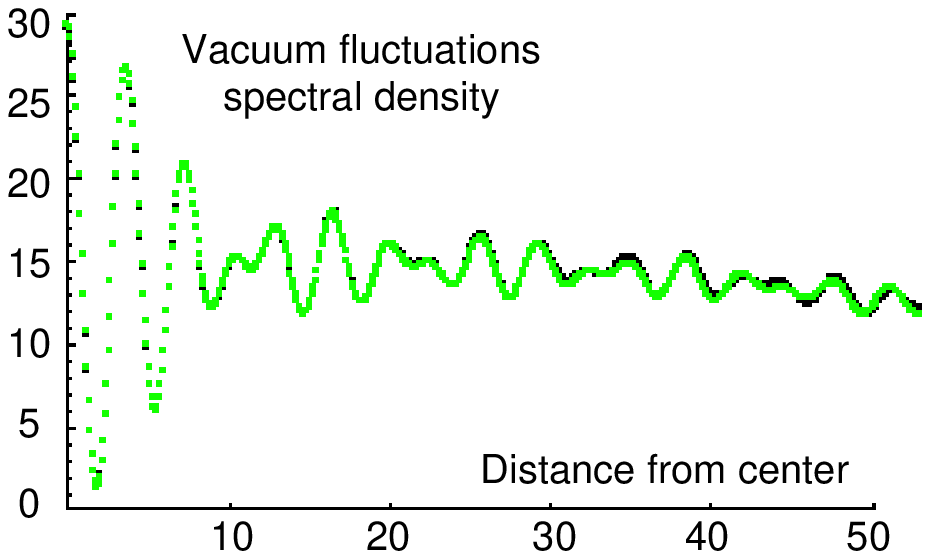} 
%\end{center}
%\vskip 5cm
\caption{ 
Same as fig. \ref{compa}, now including both spherical aberrations 
and diffraction losses.
The result is now in very good agreement with the full calculation (fig. \ref{num}),
which is also shown for comparison. }
 \label{compc}
\end{figure}

\paragraphe{Focusing defects}
The numerical scheme of the previous sections allows us to systematically study
the effects of mechanical defects in the cavity, e.g. defocusing, 
or non perfectly spherical mirrors. For instance, one can reproduce 
an axial mispositioning of the mirrors by a length $\delta$, 
by adding an imaginary part to the mirror reflectivity :
$\rho(\theta) = \rho_0 e^{2 i k \sqrt{R^2 + \delta^2 + 2 R \delta \cos\theta} -R} =
\rho_0 e^{2 i k \delta \cos\theta}$. 
The first effect of such a mispositioning is to shift the resonance frequency.
%, as it can be seen on Figure.
Correcting for this shift, the second effect is to decrease the enhancement effect,
which is typically halved for $\delta = 40 nm (k\delta = 0.3)$ with the previous
parameters. Here again, it can be seen that the ray formula gives the right answer.
As it was discussed above for positions outside the cavity center,
the main feature is indeed the phase shift after one reflection, which is correctly
described by the modified value of $\rho(\theta)$, 
while the tilting and non-imaging effect can be neglected.

\subsection{Polarization effects}
The above considerations, which were done for a scalar field, can
be straightforwardly extended to the case of a vector field,
which is needed to describe polarization effects. 
Here we will skip the explicit operatorial formulas,
and give only the results obtained in the ray optics approximation. 
As previously, this approximate solution was checked by comparison
with the complete numerical calculation, and found to be in complete agreement with it. 
For a transverse vector field $\bf \phi(\rr)$ and for two polarization directions
${\bf \epsilon_1}$ and ${\bf \epsilon_2}$, 
the results obtained in the scalar case (eq. \ref{rsr}) are then changed into~:
\ben
\label{rsrpol}
{ \ew{\bf (\epsilon_1.\phi(\rr))^* \times (\epsilon_2.\phi(\rr))}_{_{\scriptstyle \rm cav}}\over
\ew{\bf (\epsilon_1.\phi(\rr))^* \times (\epsilon_1.\phi(\rr))}_{_{\scriptstyle \rm vac}} } = 
\int {d\dir\over 4\pi} \; \frac{3}{2} \left( {\bf \epsilon_1. \epsilon_2 - (\dir.\epsilon_1)(\dir. \epsilon_2)} \right)
\times \nonumber \\
\left( {T\over \be{1- \rho e^{2i \phi}}^2} \cos^2\bigl(k\dir.\rr) \; + \;
 {T\over \be{1+ \rho e^{2i \phi}}^2} \sin^2\bigl(k\dir.\rr) \right)
\een

From this equation, the cavity induced damping and level shifts can be
obtained using eqs. (1) and (2) by integration over the frequency, which is straightforward
for the damping, and requires contour integration for the level shifts (see Appendix B). 
Finally, the effect of the cavity can be described
to a very good approximation by the following formulas~:
\beq  \label{eq:gammafinal}
{\Gamma ({\bf r}) \over \Gamma_{vac}}=
\int_{4 \pi} {d\dir\over 4\pi} \; \;{3\over 2}  (1 - ({{\bf d}.\dir \over d})^2) 
\left( {T\over \be{1- \rho e^{2i \phi}}^2 }\cos^2\bigl(k\dir.\rr) \; +\;
{T\over \be{1+ \rho e^{2i \phi}}^2} \sin^2\bigl(k\dir.\rr) \right)
\eeq
\beq \label{eq:deltafinal}
{\Delta' ({\bf r}) \over \Gamma_{vac}}=
\int_{4 \pi} {d\dir\over 4\pi}\; \;{3\over 2}  (1 - ({{\bf d}.\dir \over d})^2) 
\left({\rho \; \sin(2 \phi)\over \be{1- \rho e^{2i \phi}}^2 }\cos^2\bigl(k\dir.\rr) \; - \;
{\rho \; \sin(2 \phi)\over \be{1+ \rho e^{2i \phi}}^2} \sin^2\bigl(k\dir.\rr) \right)
\eeq
where the notation $\dir$ describes a direction in space, while $\phi$ is 
a cavity detuning parameter that will be detailed below. 
As previously, these expressions have
a straighforward interpretation, because they appear basically as integrals over the 
direction of light rays : in the integral over the directions,
$\rho$ is the mirror reflectivity for rays subtended by the cavity, and 
is zero for rays outside the cavity solid angle. The different factors 
appearing in the integrals are detailed below.

The first factor under the integral
corresponds to polarisation effects, taking into account the transverse
character of the field. 

The second (resonance) factor is of the usual Fabry-Perot
form, where $\phi$ is the cavity phase shift which includes first
a term $\phi_0 = \omega_0 R/c$. As it was shown before,
in order to obtain a correct result outside the cavity center, $\phi$ must
include also a contribution from spherical aberrations, that is :
$\phi = \phi_0 + \frac{k (r^2 - (\dir.\rr)^2)}{2 R}$. This second term 
corresponds to the extra phase shift experienced by rays
going through point $\rr$ while  propagating  along the $\dir$ direction.
The resonance factor has obviously different expressions
for the damping and the lamb shift
terms, which correspond respectively to the active and reactive parts
of the coupling. This is clearly apparent from the integrals of eqs.1 and 2, 
which involve either a delta function or a principal part. 
In the first case, the integration is trivial, and yields
 the resonance term of eq. \ref{eq:gammafinal}, while in the second case
the result is obtained by contour integration, and gives the ``dispersive" 
second term of eq. \ref{eq:deltafinal}.

The third term under the integrals 
is the stationnary wave pattern corresponding either 
to odd modes (which have an anti-node in the center and a $\cos^2\bigl(k\dir.\rr)$
space dependence) or to even modes (which have a node in the center and a 
$\sin^2\bigl(k\dir.\rr)$ space dependence). 

Finally, the integration over the mirrors is conveniently performed in spherical 
coordinates, by taking the $z$ axis along the cavity axis, and varying the 
azimuthal angle $\theta$ from $0$ to $\theta_{mirror}=\theta_{m}$. 
Improved accuracy (better than $1 \%$) is obtained
if one takes into account the fact that the rays which would 
be reflected near the edge of the mirror are actually lost due to diffraction
and fail to do as many round-trips as the other ones. As before, 
this effect can be taken into account very simply 
by decreasing $\theta_{m}$ to
$\theta_{eff} = \theta_{m} - \delta \theta$, with 
$\delta \theta = {1 \over \sqrt{k R T}}$ for symmetrical mirrors. 

The first results which can be obtained from the previous formulas are obviously
the shift and damping at the cavity center, as a function of the atom-cavity
detuning. For a dipole orientation parallel to the cavity axis, we obtain 
straightforwardly :
\beq
{\Gamma_{par} ({\bf 0}) \over \Gamma_{vac}}= 
{\Delta \Omega_{vac} \over 4 \pi} (1 + {\sin^2\theta_m \over 2})+
{\Delta \Omega_{cav} \over 4 \pi} (1 - {\cos\theta_m (1+\cos\theta_m) \over 2})  
 {T\over \be{1 - \rho e^{2i \phi_0}}^2 }
\eeq
\beq
{\Delta'_{par} ({\bf 0}) \over \Gamma_{vac}}=
{\Delta \Omega_{cav} \over 4 \pi} (1 - {\cos\theta_m (1+\cos\theta_m) \over 2})  
{\rho \; \sin(2 \phi_0)\over \be{1 - \rho e^{2i \phi_0}}^2 }
\eeq
while for a dipole orientation perpendicular to the cavity axis, we have :
\beq
{\Gamma_{perp} ({\bf 0}) \over \Gamma_{vac}}= 
{\Delta \Omega_{vac} \over 4 \pi} (1 - {\sin^2\theta_m \over 4})+
{\Delta \Omega_{cav} \over 4 \pi} (1 + {\cos\theta_m (1+\cos\theta_m) \over 4})  
 {T\over \be{1 - \rho e^{2i \phi_0}}^2 }
\eeq
\beq
{\Delta'_{perp} ({\bf 0}) \over \Gamma_{vac}}=
{\Delta \Omega_{cav} \over 4 \pi} (1 + {\cos\theta_m (1+\cos\theta_m) \over 4})  
{\rho \; \sin(2 \phi_0)\over \be{1 - \rho e^{2i \phi_0}}^2 }
\eeq
We note that these expressions yield for a randomly oriented dipole :
\beq
{\Gamma_{av} ({\bf 0}) \over \Gamma_{vac}}= 
{\Delta \Omega_{vac} \over 4 \pi}+
{\Delta \Omega_{cav} \over 4 \pi} {T\over \be{1 - \rho e^{2i \phi_0}}^2 },\;\;\;
{\Delta' ({\bf 0}) \over \Gamma_{vac}}=
{\Delta \Omega_{cav} \over 4 \pi} 
{\rho \; \sin(2 \phi_0)\over \be{1 - \rho e^{2i \phi_0}}^2 }
\eeq
corresponding to the scalar case already given above.
We note that these results are the same as those given in ref.\cite{HCTF}, up to factor two
resulting from the fact that this reference was considering spatially averaged
values rather than the peak value at the cavity center (see below for the space dependence).
These functions are plotted on fig. \ref{ad} for ${\Omega_{cav} \over 4 \pi} = 0.3$ and 
$\rho = 0.98$. It can be seen that very significant effects
occur for these quite reasonable parameters, yielding more than 30-fold increase 
in the damping rate at the cavity center. 

\begin{figure}[ht]
%\begin{center}  
\includegraphics[width=7cm]{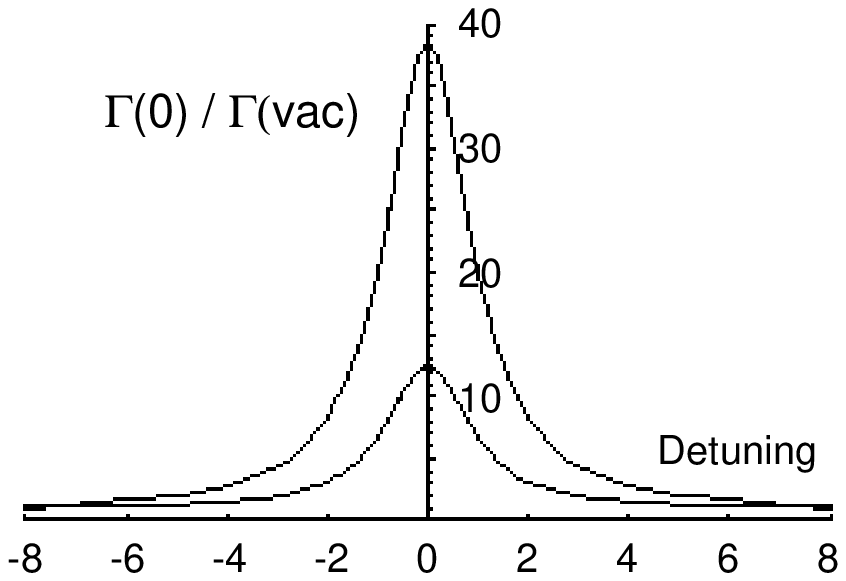} 
\includegraphics[width=7cm]{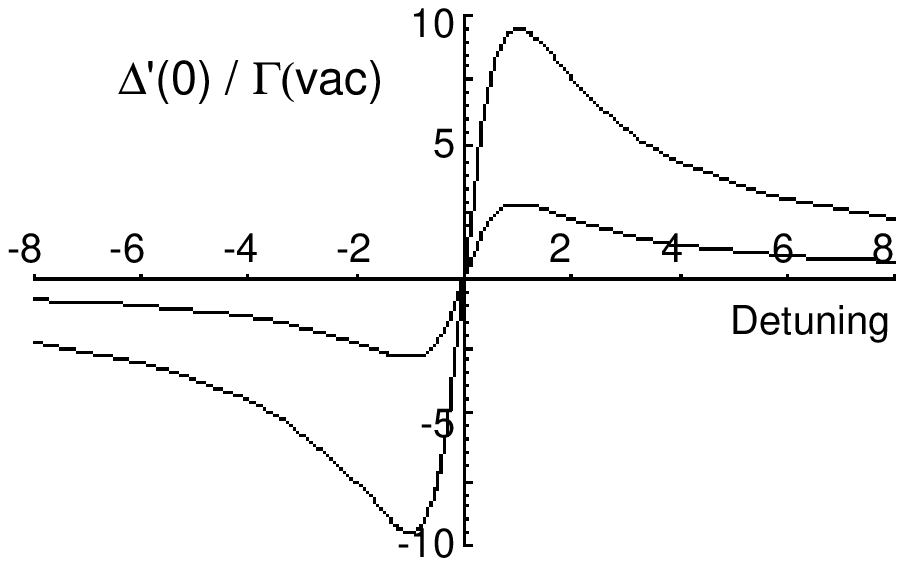} 
%\end{center}
%\vskip 5cm
\caption{ 
Normalized damping $\Gamma(0) /\Gamma_{vac}$ (left) and level shift 
$\Delta'(0) /\Gamma_{vac}$ (right) at the cavity center, as a function of 
the atom-cavity detuning normalized to the cavity linewidth. 
The amplitude reflection coefficient of the mirrors is taken to be $\rho=0.98$,
and the numerical aperture of the cavity is 0.7.
The upper curves correspond to  a dipole oriented perpendicular to the cavity axis, 
and the lower curves to a dipole oriented along the cavity axis.}
 \label{ad}
\end{figure}

The above formulas also give the damping and level shifts
as a function of space for a given frequency,
which are an important result of the present paper. 
The results in the most general case where the two mirrors have different reflectivities
are given in Appendix C. 
Two  atom-cavity detunings are specially worth looking at : 
the resonant frequency at the cavity center,
which yields maximum change in the damping rate but no cavity shift, and  frequencies
detuned by plus or minus
 half a cavity linewidth, which yield maximum cavity shifts.
These results will be exploited in the following paper \cite{JMD2}, which 
deals with vacuum-induced light forces acting on an atom close to the cavity center.

\section{Conclusion}

As a conclusion, we have calculated explicitly the cavity induced damping and 
level shifts for an atomic dipole close to the center of a spherical cavity. 
Our results are valid for arbitrary (large) aperture and (not too large) mirrors reflectivities
(for the most general case see Appendix C). 
These results show that macroscopic cavities with large 
numerical apertures are interesting candidates for cavity QED experiments in 
the optical domain. 
In particular, we show in a joint paper \cite{JMD2} that the cavity-induced
level shifts are responsible for a ``vacuum-field" force on
at atom moving close to the cavity center \cite{HBR,ESBS}. An explicit expression of the 
trapping potential can be obtained from the results given above. 

\section*{Appendix A}

When the mirrors'transmission are different, eq. \ref{rsr}
can be generalized to~: 
\beq
\label{rsr2}
{ \ew{\be{\phi(\rr)}^2}_{_{\scriptstyle \rm cav}}\over
\ew{\be{\phi(\rr)}^2}_{_{\scriptstyle \rm vac}} } = 
\int {d\dir\over 4\pi} \; M({\bf r}, 
\phi = k R + {k (r^2 - (\dir.\rr)^2) \over 2 R})
\eeq
where $r = |\; {\bf r}\; |$, and : 
\ben
M = \frac{{{\tau_1}^2}\,\left( 1 + {{\rho_2}^2} +  2\,\rho_2\,\cos (2\,
           ( \phi - k \dir . \rr ) ) \right)} 
   {2\,\left( 1 - {e^{-4\,i\, \phi}}\,\rho_1\, \rho_2 \right) \,
     \left( 1 - {e^{4\,i\, \phi}}\,\rho_1\,\rho_2 \right)}&    +  \nonumber \\
 \frac{{{\tau_2}^2}\,\left( 1 + {{\rho_1}^2} +  2\,\rho_1\,\cos (2\,
            ( \phi + k \dir . \rr ) ) \right) } 
   {2\,\left( 1 - {e^{-4\,i\, \phi}}\,\rho_1\, \rho_2 \right) \,
     \left( 1 - {e^{4\,i\, \phi}}\,\rho_1\,\rho_2 \right)  }&
 \label{deff}
\een
where for each mirror $\tau_i^2+\rho_i^2 = 1$. This equation can also 
be written in the less compact but more transparent form :
\ben
M = \frac{( 1 - \rho_1\,\rho_2 ) \,
      \left( 1 + \rho_1\,\rho_2 +  ( \rho_1 + \rho_2  ) \,
         \cos (2\, \phi) \right)} 
   {\left( 1 - {e^{-4\,i\, \phi}}\,\rho_1\, \rho_2 \right) \,
     \left( 1 - {e^{4\,i\, \phi}}\,\rho_1\,\rho_2 \right)} \;{{\cos^2 (k \dir . \rr)}}&    +  \nonumber \\
\frac{( 1 - \rho_1\,\rho_2 ) \,
      \left( 1 + \rho_1\,\rho_2 -  ( \rho_1 + \rho_2  ) \,
         \cos (2\, \phi) \right)} 
   {\left( 1 - {e^{-4\,i\, \phi}}\,\rho_1\, \rho_2 \right) \,
     \left( 1 - {e^{4\,i\, \phi}}\,\rho_1\,\rho_2 \right)}  \;{{\sin^2 (k \dir . \rr)}}&    +  \nonumber \\
 \frac{ ( 1 + \rho_1\,\rho_2 ) \,( \rho_2 -\rho_1 ) \,
      \sin (2\, \phi)} 
   {\left( 1 - {e^{-4\,i\, \phi}}\,\rho_1\, \rho_2 \right) \,
     \left( 1 - {e^{4\,i\, \phi}}\,\rho_1\,\rho_2 \right)  } \; \sin (2 k \dir . \rr)&
\een
which has the same interpretation as eq. \ref{rsr} : the 
$\sin^2 (k \dir . \rr)$ and   $\cos^2 (k \dir . \rr)$ correspond to the contributions
of the in-phase and out-of-phase standing waves along 
the direction $\dir$, while $\sin (2 k \dir . \rr) =
2 \sin (k \dir . \rr) \cos (k \dir . \rr)$
is an interference term due to the intensity inbalance between the
forward and backward contributions. 

In the case of a high finesse cavity ($\tau_1, \; \tau_2 <<1$),
these equations can be rewritten :
\beq
M = \frac{ 2\,\left(  \tau_1^2 \, \cos^2 (\phi - k \dir . \rr)  + 
                  \tau_2^2 \, \cos^2 (\phi + k \dir . \rr)  \right) }
   { \left( e^{4\,i\, \phi} -1 + (  \tau_1^2 + \tau_2^2  ) / 2 \right) \,
     \left( e^{-4\,i\, \phi}-1 + (  \tau_1^2 + \tau_2^2  ) / 2 \right) }
\eeq
or alternatively :
\beq
M = \frac{2 \, \left( \tau_1^2 + \tau_2^2 \right) \,
      \left( \cos^2 (\phi) \, \cos^2 (k \dir . \rr) + 
             \sin^2 (\phi) \, \sin^2 (k \dir . \rr) \right) \, + \,
      \left( \tau_1^2 - \tau_2^2 \right)
 \sin (2\, \phi)\,\sin (2\,k \dir . \rr)}
   { \left( e^{4\,i\, \phi} -1 + (  \tau_1^2 + \tau_2^2  ) / 2 \right) \,
     \left( e^{-4\,i\, \phi}-1 + (  \tau_1^2 + \tau_2^2  ) / 2 \right) }
\eeq
For a symetrical high-finesse cavity with $\tau_1=\tau_2=\tau$, one obtains finally :
\beq
M = \frac{4 \, \tau^2 \,
      \left( \cos^2 (\phi) \, \cos^2 (k \dir . \rr) + 
             \sin^2 (\phi) \, \sin^2 (k \dir . \rr) \right)}
   { ( e^{4\,i\, \phi} -1 + \tau^2) \,( e^{-4\,i\, \phi}-1 + \tau^2) }
\eeq
which can also be obtained directly from eq. \ref{rsr}.

\section*{Appendix B}

Let us consider the normalized Airy function~:
\beq 
{\cal L}(\phi) = \frac{\sqrt{1+F}}{1 + F \sin^2 \phi} = \frac{1-\rho^2}{| 1 - \rho e^{2 i \phi}|^2} 
\eeq 
where $F$ is related to the mirrors amplitude reflectivity by $F = 4 \rho/(1-\rho)^2$. 
In order to calculate the level shift, we need to evaluate  the principal part integral~:
\beq
\Delta(\phi) =  \int {\cal P}\frac{d\delta}{\delta}  \;  {\cal L}(\phi - \delta) 
\eeq
Replacing ${\cal L}(\phi - \delta) $ by its uneven part 
$\frac{1}{2}({\cal L}(\phi -\delta) - {\cal L}(\phi + \delta) )$, $\Delta(\phi)$ can be expressed as a standard integral~:
\beq
\Delta(\phi) = \frac{1}{2} \int \frac{d\delta}{\delta} 
(\frac{\sqrt{1+F}}{1 + F \sin^2( \delta - \phi)} - \frac{\sqrt{1+F}}{1 + F \sin^2(\delta +\phi)})
\eeq
This quantity can be evaluated by contour integration, using the zeros of
the denominators $1 + F \sin^2( \delta \pm \phi)$, which are respectively $\delta_n^- = - \phi \pm i \beta + n \pi$, 
and $\delta_n^+ = \phi \pm i \beta + n \pi$, where $\beta >0$ and $\sinh^2 \beta = 1/F$. 
Using a contour in the lower part of the complex plane, which includes
the poles $\delta_n^\pm = \pm \phi - i \beta + n \pi$ with $n = ... -1, 0, 1, ...$, we obtain for instance~:
\beq
\int \frac{d\delta}{\delta} \frac{1}{1 + F \sin^2(\delta-\phi )} =
\frac{-  2 i \pi }{F \sin(-2 i \beta)} \; \Sigma_n \frac{1}{\phi - i \beta + n \pi} =
\frac{2 \pi}{F \sinh(2 \beta)}  \frac{1}{\tan(\phi - i \beta)} 
\eeq
Using $\sinh(2 \beta) = 2 \sqrt{1+F}$ and  $\tan(\phi - i \beta) = 
\frac{\sqrt{1+F} \sin \phi - i \cos\phi}{\sqrt{1+F} \cos \phi + i \sin \phi}$, we obtain thus~:
\beq
\Delta(\phi) = \frac{\pi}{2} \left( \frac{1}{\tan(\phi - i \beta)}  - \frac{1}{\tan(-\phi - i \beta)} \right) =
\frac{\pi F}{2} \left( \frac{\sin(2 \phi)}{1 + F \sin^2 \phi} \right)
\eeq
Coming back to mirrors reflectivities, we obtain finally the expression used in eq. \ref{eq:deltafinal}~:
\beq
\Delta(\phi) = 2 \pi  \; \frac{\rho \sin(2 \phi)}{| 1 - \rho e^{2 i \phi} |^2}.
\eeq

\section*{Appendix C}

In the case of asymetrical mirrors with amplitude transmitivities $\rho_1$ and $\rho_2$, 
one can use the formulas of Appendix B, changing $F$ into $F' = 4 \rho_1 \rho_2 /(1-\rho_1 \rho_2)^2$,
and $\phi$ into $\phi' = 2 \phi$, so that~:
\beq
\Delta(\phi') = \frac{\pi \; F'}{2} \left( \frac{\sin(2 \phi')}{1 + F' \sin^2 \phi'} \right)
\eeq
In addition, we need to evaluate the principal parts for ${\cal L}(\phi - \delta)$ multiplied
either by  $\cos(\phi - \delta)$ or $\sin(\phi - \delta) $. 
Taking for instance the cosine part, we obtain~:
\beq
\Delta_c(\phi) =  \int {\cal P}\frac{d\delta}{\delta}  \;  {\cal L}(\phi - \delta) \; \cos(\phi - \delta)
\eeq
This can be done as before, and we have~:
\beq
\int \frac{d\delta}{\delta} \frac{\cos( \delta - \phi)}{1 + F \sin^2(\delta-\phi )} =
\frac{-  2 i \pi }{F \sin(-2 i \beta)} \; \Sigma_n \frac{\cos(-i\beta + n \pi) }{\phi - i \beta + n \pi} =
\frac{2 \pi \; \cosh(\beta )}{F \sinh(2 \beta)}  \; \frac{1 }{\sin(\phi - i \beta)}
\eeq
Using $\sin(\phi - i \beta) = (\sqrt{1+F} \sin \phi - i \cos\phi)/\sqrt{F}$ we get :
\beq
\Delta_c(\phi) =  \frac{\pi \; \sqrt{F(1+F)}}{2 F }  \; 
(\frac{1 }{\sin(\phi - i \beta)} - \frac{1 }{\sin(-\phi - i \beta)}) =
\pi \; (1+F) \frac{\sin \phi}{ 1 + F \sin^2 \phi}.
\eeq
Applying the same method for the sine part, we obtain finally~:
\beq
\Delta_c(\phi') =  \pi \; (1+F') \frac{\sin \phi'}{ 1 + F' \sin^2 \phi'} \; \; \; \; 
\Delta_s(\phi') =  -\pi \; \frac{\cos \phi'}{ 1 + F' \sin^2 \phi'}.
\eeq
From these formulas, we obtain the damping and level shift in the asymetrical case~:
\ben  \label{eq:gammafinalas}
{\Gamma ({\bf r}) \over \Gamma_{vac}}=
\int_{4 \pi} {d\dir\over 4\pi} \; \;{3\over 2}  (1 - ({{\bf d}.\dir \over d})^2) \{  %\nonumber \\
&&\frac{( 1 - \rho_1\,\rho_2 ) \,
      \left( 1 + \rho_1\,\rho_2 +  ( \rho_1 + \rho_2  ) \,
         \cos (2\, \phi) \right)} 
   {\left( 1 - {e^{-4\,i\, \phi}}\,\rho_1\, \rho_2 \right) \,
     \left( 1 - {e^{4\,i\, \phi}}\,\rho_1\,\rho_2 \right)} \;{{\cos^2 (k \dir . \rr)}}    +  \nonumber \\
&&\frac{( 1 - \rho_1\,\rho_2 ) \,
      \left( 1 + \rho_1\,\rho_2 -  ( \rho_1 + \rho_2  ) \,
         \cos (2\, \phi) \right)} 
   {\left( 1 - {e^{-4\,i\, \phi}}\,\rho_1\, \rho_2 \right) \,
     \left( 1 - {e^{4\,i\, \phi}}\,\rho_1\,\rho_2 \right)}  \;{{\sin^2 (k \dir . \rr)}}    +  \nonumber \\
&& \frac{ ( 1 + \rho_1\,\rho_2 ) \,( \rho_2 -\rho_1 ) \,
      \sin (2\, \phi)} 
   {\left( 1 - {e^{-4\,i\, \phi}}\,\rho_1\, \rho_2 \right) \,
     \left( 1 - {e^{4\,i\, \phi}}\,\rho_1\,\rho_2 \right)  } \; \sin (2 k \dir . \rr) \}
\een
\ben \label{eq:deltafinalas}
{\Delta' ({\bf r}) \over \Gamma_{vac}}=
\int_{4 \pi} {d\dir\over 4\pi} && {3\over 2}  (1 - ({{\bf d}.\dir \over d})^2)   \; \{  %\nonumber \\
\frac{ \rho_1\,\rho_2 \, \sin(4 \phi) + ( \rho_1 + \rho_2  )( 1 + \rho_1\,\rho_2) \, \sin(2 \phi)/2 } 
   {\left( 1 - {e^{-4\,i\, \phi}}\,\rho_1\, \rho_2 \right) \,
     \left( 1 - {e^{4\,i\, \phi}}\,\rho_1\,\rho_2 \right)} \;{{\cos^2 (k \dir . \rr)}}    +  \nonumber \\
&&\frac{\rho_1\,\rho_2 \, \sin(4 \phi) - ( \rho_1 + \rho_2  )( 1 + \rho_1\,\rho_2) \, \sin(2 \phi)/2 } 
   {\left( 1 - {e^{-4\,i\, \phi}}\,\rho_1\, \rho_2 \right) \,
     \left( 1 - {e^{4\,i\, \phi}}\,\rho_1\,\rho_2 \right)}  \;{{\sin^2 (k \dir . \rr)}}  +  \nonumber \\
&& \frac{ ( 1 - \rho_1\,\rho_2 ) \,( \rho_1 -\rho_2 ) \,
      \cos (2\, \phi)/2} 
   {\left( 1 - {e^{-4\,i\, \phi}}\,\rho_1\, \rho_2 \right) \,
     \left( 1 - {e^{4\,i\, \phi}}\,\rho_1\,\rho_2 \right)  } \; \sin (2 k \dir . \rr)\}.
\een
An particularly interesting  case is a one-mirror cavity ($\rho_1=\rho$, $\rho_2=0$), for which
\ben  \label{eq:gammafinalaz}
{\Gamma ({\bf r}) \over \Gamma_{vac}}=
\int_{4 \pi} {d\dir\over 4\pi} \; \;{3\over 2}  (1 - ({{\bf d}.\dir \over d})^2) \; 
\{ 1 + \rho  \; \cos(2 (k \dir . \rr + \phi)) \}
\een
\ben \label{eq:deltafinalaz}
{\Delta' ({\bf r}) \over \Gamma_{vac}}=
\int_{4 \pi} {d\dir\over 4\pi}\; \;{3\over 2}  (1 - ({{\bf d}.\dir \over d})^2)  \; 
\{ \frac{\rho}{2} \; \sin(2 (k \dir . \rr + \phi) \}
\een
Neglecting spherical aberrations, one has as before $\phi = \omega_0 R/c$, 
while ${\bf r}$ corresponds to the atom's position with respect to the mirror's center of curvature. 
We note that these equations have the correct behaviour $\Gamma ({\bf r}) = \Gamma_{vac}$
and $\Delta' ({\bf r}) = 0$ if $\rho = 0$ (no cavity). 
In the case of a small solid angle $\epsilon =\Omega/(4 \pi)$ subtended by the spherical mirror
and a dipole orthogonal to the ``cavity" axis $Oz$, one gets~:
\beq
{\Gamma (z) \over \Gamma_{vac}}Ê\approx 1 + {3 \epsilon \rho \over 2} \; \cos(2 (k z + \phi)) ,
\; \; \; \; \; 
{\Delta' (z) \over \Gamma_{vac}} \approx
{3 \epsilon \rho \over 4}   \; \sin(2 (k z +\phi)).
\label{dzz}
\eeq
From eq. \ref{deff}, the total phase $(k z +\phi)$ corresponds to the distance $l$ between the atom and mirror 1. 
Taking into account that $k z + \phi = 0$, or equivalently $kl  = \pi/2$ mod $\pi$, are antinodes of the standing wave,
one has more  precisely $kl = k z +\phi + \pi/2$.
Eq. \ref{dzz} corresponds then to the results obtained in ref. \cite{dz}, up to a factor $3/2$
due to the fact that the vectorial character of the dipole was ignored in ref. \cite{dz}.
More accurate results for any position of the atom and solid angle subtended by the mirror can be obtained from
eq. \ref{eq:gammafinalaz} and \ref{eq:deltafinalaz}.

\end{document}